\documentclass[twoside,fleqn]{article}
\usepackage{hyperon99_paper}
\usepackage{acromake}
\usepackage{axodraw}
\usepackage{cite}
\usepackage{epsf}
\usepackage{topcapt}
\usepackage{setup}
\title{%
  \hfill\raisebox{2cm}[0pt][0pt]{\normalsize EPTCO-99-003} \\
  Theory Summary Talk\thanks{Presented at Hyperon~99 Symposium, Fermilab,
  Batavia, Sept.~27--29, 1999}}
\author{Philip G. Ratcliffe
  \address{
    Dip.\ di Scienze,
    Univ.\ degli Studi dell'Insubria---sede di Como,
    via Lucini 3, 22100 Como, Italy \\
    and Istituto Nazionale di Fisica Nucleare---sezione di Milano,
    via Celoria 16, 20133 Milano, Italy \\
  }\thanks{E-mail:~\texttt{pgr@fis.unico.it}}
}
\begin{document}
\begin{abstract}
A summary is presented of the more theoretical aspects of the presentations
made at Hyperon~99. In addition, some material is covered which was not
presented at the symposium but which I feel is pertinent to the main theme of
hyperons and/or, more in particular, to discussions conducted during the
symposium.
\end{abstract}
\maketitle
\section{INTRODUCTION}
In this summary talk I shall attempt not only to highlight some of the issues
touched upon by the speakers at this symposium but also to cover some of those
topics which, for one reason or another, were left uncovered. All or most of
the topics mentioned by Holstein in his opening overview talk were indeed
dealt with during the last few days. However, some deserving issues, notably
hypernuclei, were left untouched as too were other aspects of hyperon physics
not considered by Holstein\cite{Holstein:1999x1}. One of these is the
large-$N_c$ expansion, which has been promoted in this context by the San
Diego group and which should have been aired by Liz Jenkins who unfortunately
was unable to attend, another is the possibility of using perturbative QCD to
describe large-$p_T$ semi-inclusive hyperon production and the associated
surprisingly large measured transverse hyperon polarisations. Last but not
least, is the case of hypernuclei, which was only very briefly mentioned by
Holstein.

While I hope to do justice to the speakers and, in particular, to the missing
subjects, time and space clearly do not permit as complete a job as I might
have liked. Thus, I shall only attempt to give a flavour of what was discussed
here at Fermilab and its relevance to the future of hyperon physics programmes
but leave the details to the speakers contributions; and also to fill in what
I feel were important gaps, with at least a hint of what might have been said.
Of course, for the full details the reader is referred to the original talks.

Before turning to the more serious part of the talk, the use of the expression
``perturbative QCD'' reminds of something that struck me on the flight from
Italy to Chicago. Along with the pre-packed lunch, came a salad and a small
container of what purported to be ``Creamy Italian Dressing''. Now, it might
even be that the long list of exotic ingredients adds to its appeal in the
eyes of some, and, certainly, hidden in there were the three prime ingredients
I have been taught to use in Italy: namely, oil (though it should be of the
olive variety and not soybean), vinegar and salt. However, I strongly doubt
that any Italian in the audience will ever have put such a concoction onto his
or her salad at home. That is to say: we have on occasions (albeit only a very
few) at this symposium heard the words QCD mentioned, but it would be very
hard indeed, given the usual trimmings or non-perturbative model input, to
extract anything about the presumed fundamental theory of hadronic
interactions itself from the sort of phenomenology discussed. On the one hand,
this is refreshing for those of us who are a little weary of hearing about the
latest $n$-loop or next-to-next-to\dots-leading-order calculation. On the
other, there should be a wariness that much of the model building that goes on
in hadronic physics, with the ever-comforting benefit of hindsight, often
risks being little more than a patching-up job on a rather cloudy situation.

Let me now turn to the task in hand, I have divided the summary talk into
sections describing: the static properties of hyperons; semi-leptonic,
radiative and non-leptonic hyperon decays; hyperon polarisation and
hypernuclei; with a little space dedicated to some concluding remarks.
\section{STATIC PROPERTIES}
I shall consider here the description of masses and magnetic moments (in
particular, those of the baryon octet). Lipkin\cite{Lipkin:1999x1} reminded us
of a remarkable series of predictions of the na\"{\i}ve quark model: \eg, the
relation between the baryon and pseudoscalar and spin-one meson octet masses,
\begin{equation}
  m_\Lambda - m_N
  \simeq
  \frac{3(m_{K^*}-m_K)+(m_K-m_\pi)}{4};
\end{equation}
experimentally, the left-hand side is 177\,MeV and the right-hand side,
180\,MeV. A similarly successful relation between the baryon octet and
decuplet and pseudoscalar and spin-one meson masses is provided by
\begin{equation}
  \frac{m_\Delta-m_N}{m_{\Sigma^*}-m_\Sigma}
  \simeq
  \frac{m_\rho-m_\pi}{m_{K^*}-m_K};
\end{equation}
where experimentally the two sides are 1.53 and 1.61 respectively. The simple
but nevertheless important conclusion to be drawn is that quarks bound inside
mesons behave just like quarks inside baryons.

With regard to the magnetic moments, there exist further simple relations:
\begin{equation}
  \mu_p + \mu_n
  \simeq
  \frac{2m_p}{m_N+m_\Delta},
\end{equation}
where the experimental values are 0.880 and 0.865 respectively, and
\begin{equation}
  \mu_\Lambda
  \simeq
  -\frac13 \mu_p \frac{m_{\Sigma^*}-m_\Sigma}{m_\Delta-m_N},
\end{equation}
here the experimental values are both -0.61!

Since Liz Jenkins was unable to be present at the symposium and despite
Lipkin's bold attempt at an impersonation, we did not learn anything about the
$1/N_c$-expansion approach\cite{tHooft:1974x1} and the work of the San Diego
group. It is impossible here to do justice to this field and the interested
reader is referred to the comprehensive review article by
Jenkins\cite{Jenkins:1998wy}. Let me simply try to give a flavour of what is
involved and the results achieved.

A spin-flavour symmetry is found to emerge for baryons in the large-$N_c$
limit; large-$N_c$ baryons form irreducible representations of the
spin-flavour algebra, and their static properties may be computed in a
systematic expansion in $1/N_c$. Symmetry relations for static baryon matrix
elements may then be obtained at various orders in the $1/N_c$ expansion by
neglecting sub-leading $1/N_c$ corrections. These symmetry relations (such as
those already mentioned) may then be arranged according to a $1/N_c$
hierarchy, \ie, the higher the order in $1/N_c$ is the sub-leading correction,
the better one expects the relation to be satisfied. For QCD baryons with
$N_c=3$ one then naturally expects such a hierarchy to be based on steps of
roughly 1/3.

\begin{figure}[htb]
  \epsfxsize=75mm
  \epsfbox{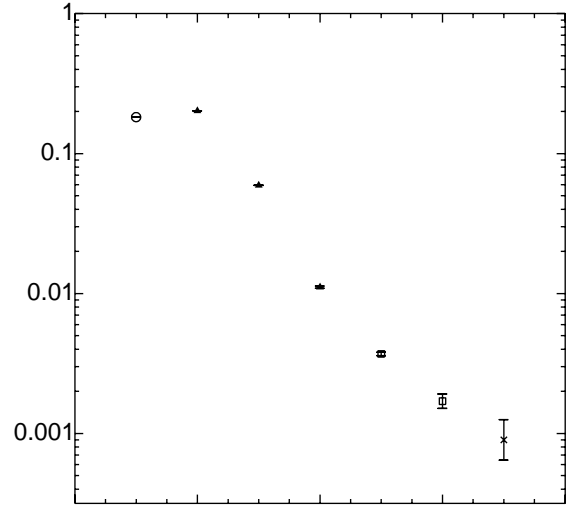}
  \caption{Isospin-averaged baryon-mass (normalised) combinations
    from\protect\cite{Jenkins:1995x1}. The error bars are experimental and the
    horizontal scale is merely a label for the given combinations. The
    open circle is an $\mathcal{O}(1/N_c^2)$ mass combination; the three solid
    triangles are $\mathcal{O}(\epsilon/N_c)$, $\mathcal{O}(\epsilon/N_c^2)$,
    and $\mathcal{O}(\epsilon/N_c^3)$ mass relations; the open squares are
    $\mathcal{O}(\epsilon^2/N_c^2)$, and $\mathcal{O}(\epsilon^2/N_c^3)$
    relations; and the cross is an $\mathcal{O}(\epsilon^3/N_c^3)$ relation.}
  \label{fig:largenc}
\end{figure}

Thus, for example, the celebrated Coleman-Glashow mass relation,
\begin{equation}
  (p - n) - (\Sigma^+ - \Sigma^-) + (\Xi^0 - \Xi^-) = 0,
\end{equation}
is $\mathcal{O}(1/N_c)$ in the $1/N_c$ expansion, so that the mass relation
should be more accurate than would be predicted by mere flavour-symmetry
breaking arguments alone. In fig.\,\ref{fig:largenc} the diagram, taken from
Jenkins and Lebed\cite{Jenkins:1995x1}, shows a hierarchy of baryon-mass
relations in both $1/N_c$ and an $SU(3)$ flavour-symmetry breaking parameter,
$\epsilon\sim0.3$, as predicted by the theoretical analysis. The accuracy with
which the magnitude of the deviations follows the expected pattern is
striking.

Analogously, the baryon magnetic moments may also be studied. Results show
that in the large-$N_c$ limit the isovector baryon magnetic moments are
determined up to a correction of relative order $1/N_c^2$, so that the ratios
of the isovector magnetic moments are determined for $N_f=2$ flavours up to a
correction of relative order $1/N_c^2$. And again one finds that the general
$1/N_c$ hierarchy is respected.
\section{SEMI-LEPTONIC DECAYS}
\begin{figure}[htb]
  \centering
\begin{picture}(140,140)(-70,-70)
  \SetWidth{1.0}
  \SetScale{0.6}
  \ArrowLine    (-113,  15)( -68,  85)      
  \ArrowLine    (  15,  35)(  45,  85)      
  \ArrowLine    ( -65, -85)( -15,   0)      
  \ArrowLine    ( -40, 100)(  40, 100)      
  \DashArrowLine( -90,   0)( -20,  20){15}  
  \DashArrowLine(  90,   0)(  20,  20){15}  
  \DashArrowLine( -45, -85)( -15, -35){3}   
  \put          (  36, -49){.}
  \put          (  39, -44){.}
  \put          (  42, -38){.}
  \ArrowLine    (  77, -55)(  78, -53)
  \put          (  48, -28){.}
  \put          (  51, -22){.}
  \put          (  54, -17){.}
  \put          (  57, -11){.}
  \Text         ( -35,  60)[c]{$n$}
  \Text         (  35,  60)[c]{$p$}
  \Text         ( -69,   0)[c]{$\Sigma^-$}
  \Text         (   0,  12)[c]{$\Lambda^0$}
  \Text         (   0, -12)[c]{$\Sigma^0$}
  \Text         (  69,   0)[c]{$\Sigma^+$}
  \Text         ( -35, -60)[c]{$\Xi^-$}
  \Text         (  35, -60)[c]{$\Xi^0$}
\end{picture}
  \caption{The SU(3) scheme of the measured baryon-octet $\beta$-decays: the
    solid lines represent decays for which both rates and asymmetry
    measurements are available; the long dash, only rates; the short dash,
    $f_1=0$ decays; and the dotted line, the recent KTeV data.}
  \label{fig:hsdexpt}
\end{figure}
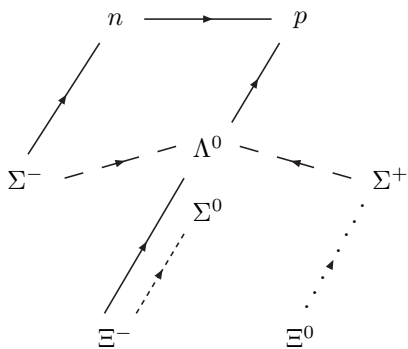

Another problem to which the $1/N_c$ expansion has been applied is that of
\HSD. Moreover at this symposium members of the KTeV collaboration have
presented their results for the hitherto completely unexplored $\XO$
$\beta$-decay, $\XO\to\SP{e}\bar\nu$. Let us first examine the experimental
situation: in fig.\,\ref{fig:hsdexpt} the measured decay modes and nature of
the data available are indicated, and in table\,\ref{tab:hsdexpt} the values
obtained for the decay widths and angular asymmetries are displayed.

\begin{table*}[htb]
  \caption{The present world \protect\HSD rate and angular-correlation data
    \protect\cite{PDG:1998x1}. The numerical values marked $g_1/f_1$ are those
    extracted from angular correlations.}
  \def\arraystretch{1.1}
  \doublerulesep=0.2pt
  \setlength{\tabcolsep}{1.5pc}
  \catcode`!=\active \def!{\hphantom{0}}
  \catcode`?=\active \def?{\hphantom{-}}
  \def\F#1{\hbox to0pt{$\;^{#1}$\hss}}
\begin{tabular*}{\textwidth}{r@{$\,\to\,$}l@{\extracolsep{\fill}}cccl}
  \hline\hline
  \multicolumn{2}{c}{Decay} &
  \multicolumn{2}{c}{Rate ($10^6$\,s$^{-1}$)} & $g_1/f_1$ & $g_1/f_1$ \\
  \cline{3-4}
  $A$   & $B\ell\nu\quad$ & $\ell=e^\pm$ & $\ell=\mu^-$
  & $\ell=e^-$  & SU(3) \\
  \hline
  $n$   & $p$   & $1.1274\pm0.0025\F{a}$ &
  & $?1.2601\pm0.0025$ & $F+D$ \\
  $\LO$ & $p$   & $3.161!\pm0.058!$      & $0.60\pm0.13$
  & $?0.718!\pm0.015!$ & $F+D/3$ \\
  $\SM$ & $n$   & $6.88!!\pm0.23!!$      & $3.04\pm0.27$
  & $-0.340!\pm0.017!$ & $F-D$ \\
  $\SM$ & $\LO$ & $0.387!\pm0.018!$      &
  & & $-\sqrt{2/3}\,D$\F{b} \\
  $\SP$ & $\LO$ & $0.250!\pm0.063!$      &
  & & $-\sqrt{2/3}\,D$\F{b} \\
  $\XM$ & $\LO$ & $3.35!!\pm0.37\F{c}!!$ & $2.1!\pm2.1\F{d}!$
  & $?0.25!!\pm0.05!!$ & $F-D/3$ \\
  $\XM$ & $\SO$ & $0.53!!\pm0.10!!$      &
  & & $F+D$ \\
  \hline\hline
\end{tabular*}
{\small
  $^a$\,Rate given in units of $10^{-3}\,$s$^{-1}$.
  $^b$\,Absolute expression for $g_1$ given ($f_1=0$).
  $^c$\,Scale factor 2 included in error (PDG practice for discrepant data).
  $^d$\,Data not used in these fits.}
\label{tab:hsdexpt}
\end{table*}

Noticeably missing from the data table is the recently published preliminary
KTeV measurement of the decay $\XO\to\SP{e}\bar\nu$, the interested reader is
referred to the talks presented here by Alavi-Harati\cite{AlaviHarati:1999x1}
and Bright\cite{Bright:1999x1}. The interesting point here is that the various
approaches to dealing with SU(3) breaking in this sector provide well-defined
and strongly bound predictions for both the decay rate and axial form factor;
the data are now tantalisingly close to differentiating between the various
predictions. It must be said (to the authors chagrin), however, that the
present world data actually still marginally favour the original SU(3)-based
prediction of Cabibbo theory\cite{Cabibbo:1963x1} (more than a third of a
century old).

A discussion of the theoretical problems involved was presented by
Garc\'{\i}a\cite{Garcia:1999x1}. While pointing out that the only severe
discrepancy (\ie, larger than three standard deviations) with respect to
Cabibbo theory lies with the rate for $\SM\to\LO{e}\nu$, Garc\'{\i}a also stressed
that the to exploit the experimental information to the full, one should fit
to the asymmetry parameters ($\alpha_{e\nu}$, $\alpha_e$, $\alpha_\nu$ and
$\alpha_{B_f}$) and not merely to the extracted value of $g_1$ alone.

\subsection{\boldmath{$F$} and \boldmath{$D$} or \boldmath{$g_A$}'s (a rose by any
  other name \dots)}
Besides the obvious possibility as a measurement of the \CKM matrix element
$V_{us}$, the solidity of which has been cast into doubt (see, \eg,
\cite{PDG:1998x1}), the study of \HSD provides unique access to the $F$ and
$D$ parameters necessary for a complete analysis of polarised \DIS. Let me
recall briefly the proton-spin story. Longitudinally polarised \DIS is
governed by the structure function $g_1(x,Q^2)$, whose integral in $x$ (the
Bjorken scaling variable or partonic momentum fraction) is given in terms of
quark spin contributions to the nucleon:
\begin{eqnarray}
  \Gamma_1^p
  &=&
  \int_0^1 \mathrm{d} \, g_1(x,Q^2)
  \nonumber
\\
  &=&
  \smfrac12 \left[
    \smfrac49\Delta{u}
  + \smfrac19\Delta{d}
  + \smfrac19\Delta{s} \right]
  \nonumber
\\
  &&
  \quad \mbox{} \times
  \left[ \vphantom{\smfrac12} 1 + \delta_\mathrm{PQCD} + \dots \right]
\end{eqnarray}
(note that here and in what follows the representation of the radiative
corrections \etc\ is only intended to be schematic). The SMC experiment, for
example, measures\cite{Adams:1994x1}
\begin{equation}
  \Gamma_1^p
  =
  0.120 \pm 0.005 \pm 0.006,
\end{equation}
which, combined with the prediction that
\begin{equation}
  \Gamma_1^p
  =
  \left[ F - \smfrac13 D + \smfrac23 \Delta{s} \right]
  \left[ \vphantom{\smfrac12} 1 + \delta_\mathrm{PQCD} + \dots \right],
\end{equation}
leads to an extracted value for the strange-quark spin $\Delta{s}\simeq-0.1$
(using $g_A^n=1.267$ and $F/D=0.58$), which is a surprisingly large value for
a sea contribution and constitutes the variously denominated spin ``crisis'',
``problem'' or ``puzzle''. The point is that if $F/D$ were to shift to 0.5
say, then the extracted value would become $\Delta{s}\simeq0$, neatly
resolving all conflict.

Given the obviously important r\^{o}le that \HSD plays in this analysis, it is
clearly vital to understand to what extent the values of $F$ and $D$ extracted
from \HSD are to be considered reliable and, perhaps more to the point, just
what they are. Thus, I would respond to Lipkin's earlier comments by saying
that ``playing'' with parametrisations (of SU(3) breaking) \emph{is}
legitimate in the context of attempting to understand what may be happening,
in order to place (reliable) bounds on other predictions. And certainly, it is
only a cosmetic question whether to parametrise using $F$ and $D$ or $g_A$ or
any other description that may have physical meaning in a given analysis.

\subsection{\boldmath{$V_{us}$} or \boldmath{$\sin\theta_\mathbf{C}$}}
In the context of this symposium probably the more interesting aspect of \HSD
is the possibility of measuring $V_{us}$. It was noted during one of the talks
that the Particle Data Group no longer considers \HSD as a reliable source of
this Standard Model parameter, preferring the so-called $K_{e3}$-decay
data\cite{PDG:1998x1}. Let me note in passing that the proton-spin analysis
does not yet require the same level of precision.

There are several difficulties that render the extraction of $V_{us}$ from
\HSD data a delicate process. First, but not foremost in the discussion of
hadronic physics, is the continuing saga of neutron $\beta$-decay; the
discrepancies present in this sector cloud the issue of \CKM unitarity and
therefore need to be resolved before real progress can be made with regard to
$V_{us}$. An oft neglected question is that of the r\^{o}le of so-called
second-class currents. These have not yet been investigated experimentally to
any real extent, except to show that their presence could have a profound
effect on the extracted value of $g_A$, possibly even shifting the ratio $F/D$
back to its original SU(6) value.

The area where most theoretical effort has been made, and using a number of
approaches, is that of SU(3) breaking. The problem here is that most of the
analyses presented in the literature to date are highly model dependent
(indeed, often the main aim is to \emph{test} the model and not necessarily
provide a reliable analysis of parameters at all). Moreover, a severe failing
of many published analyses is that they are highly selective of the data used.
While it may make sense to examine the effect of neglecting this or that data
set, if data are discarded on the basis of apparent discrepancy with SU(3)
symmetry predictions, then the resulting bias is as obvious as it is
unacceptable.

It is evident then that to resolve these difficulties, an improved
experimental database is required: the presently available data do not
sufficiently over-constrain the system, which needs different combinations
both of the $F$ and $D$ parameters and of $\Delta{S}=0$ and $|\Delta{S}|=1$
decays, and also of both rates and angular correlations for the \emph{same}
decay modes. The KTeV data will go some way to meeting this request, providing
as it does an evaluation of $F+D$, the same combination as found in neutron
$\beta$-decay. However, there are several modes that have been measured but
not yet with sufficient precision to be of real use; attempts should be made
to improve these, not forgetting an eye towards the possibility of
second-class current contributions.

In conclusion, a few comments are in order regarding the decay mode
$\XO\to\SP{e}\bar\nu$. The two predictions that have been compared at this
symposium to the KTeV results are those of Flores-Mendieta, Jenkins and
Manohar\cite{Flores-Mendieta:1998ii}, using the above-mentioned $1/N_c$
expansion, and mine\cite{Ratcliffe:1998su}, using the centre-of-mass
corrections as proposed in\cite{Donoghue:1987x1}. I should remark that the
difference between the results of these last two papers is due in part to the
publication of new data between the two, but mainly to the large strange-quark
wave-function mismatch correction applied in the latter and not in the former
(owing to its incompatibility with the later data). Thus, for this type of
approach one finds relatively small deviations (at most a few percent) and an
overall good description of the data. As for the $1/N_c$ approach, it should
be noted that there a much larger fit was performed, including data on the
weak non-leptonic decuplet decays, which apparently have a very strong
influence and lead to very large corrections in \emph{both} sectors.

\subsection{Isospin Violation}
However, before moving on to the next section, I should like to recall Karl's
talk on isospin violation in semi-leptonic decays\cite{Karl:1999x1}, indeed
his comments could have a wider impact than just on these decays. The question
regards the possible mixing between $\LO$ and $\SO$. If isospin is conserved,
then these two particles should simply correspond to the standard SU(3)
states. If, on the other hand, the isospin SU(2) is broken (as evidently it
is, slightly), then the physically observed particles will be mixture of the
na\"{\i}ve SU(3) states. The related mixing angle is typically taken to be
$\sin\phi\simeq-0.015$.

Clearly the decays in which the effects should be most felt are those
involving both $\LO$ and $\Sigma$ hyperons. Thus, for example, the ratio of
decay widths:
\begin{eqnarray}
  R(\phi)
  &=&
  \frac{\Gamma(\SP-\LO{e^+}\nu)}{\Gamma(\SP-\LO{e^-}\bar\nu)} \\
  &=& (1-4\phi) \, R(0),
\end{eqnarray}
should be shifted by about 6\% owing to the mixing\cite{Karl:1994ie}.
Unfortunately, present experimental precision is too poor (for the $\SP$
decay) to detect such a shift. There would also be consequences for the vector
coupling in these decays, which should vanish in pure Cabibbo theory but will
be non-zero if there is $\LO$-$\SO$ mixing.
\section{WEAK RADIATIVE DECAYS}
The subject of weak radiative hyperon decays has been discussed in detail by
\.Zenczykowski\cite{Zenczykowski:1999x1}. One of the central problems here is
the apparent violation of Hara's theorem\cite{Hara:1964x1}, again in existence
for over a third of a century. The decays $B\to{B}'\gamma$ can be described in
terms of the weak Hamiltonian matrix element:
\begin{equation}
  \langle B'\gamma | \mathcal{H}_W | B \rangle
  \propto
  \bar u(p')
  \varepsilon _\mu \sigma^{\mu\nu} q_\nu
  \left( C + D\gamma_5 \right)
  u(p),
\end{equation}
where the term in $C$ is magnetic and $D$ is electric. Hara's theorem is based
on U-spin and states that $D=0$ for $B$ and $B'$ belonging to the same U-spin
multiplet. From the observation that U-spin is not badly broken, one expects
$D$ to be small (say, of order 10\%). The experimental implication is a small
asymmetry parameter:
\begin{equation}
  \alpha
  =
  \frac{2\RE C^*D}{|C|^2+|D|^2},
\end{equation}
for the decays $\SP\to{p}\gamma$ and $\XM\to\SM\gamma$. Experimentally the
former is $-0.76\pm0.08$; \ie, the theorem is almost maximally violated. Such
a large value is indeed very difficult to explain consistently.

\begin{figure}[htb]
  \centering
\begin{picture}
            (180, 70)(  0,  0)
  \SetWidth {0.6}
  \SetOffset(  0,  0)
  \Line     (  0,  0)( 75,  0)
  \Photon   ( 25,  0)( 50, 50) {3} {4}
  \Vertex   ( 25,  0){2}
  \Line     ( 47, -3)( 53,  3)
  \Line     ( 53, -3)( 47,  3)
  \SetOffset(105,  0)
  \Line     (  0,  0)( 75,  0)
  \SetOffset(130,  0)
  \Photon   ( 25,  0)( 50, 50) {3} {4}
  \Vertex   ( 25,  0){2}
  \SetOffset( 80,  0)
  \Line     ( 47, -3)( 53,  3)
  \Line     ( 53, -3)( 47,  3)
\end{picture}
  \caption{The pole diagrams contributing to hyperon radiative decays;
    the cross indicates the intermediate-state insertions.}
  \label{fig:polediag}
\end{figure}
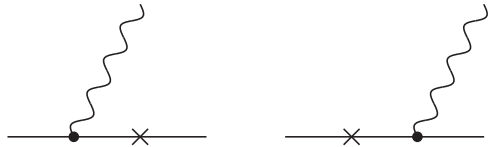

Successful approaches (salvaging Hara's theorem) may be found in the
literature, due to Le Younac \etal\cite{Leyounac:1979x1} and Borasov and
Holstein\cite{Borasov:1999x1}. The central idea of these two groups is the
insertion of additional intermediate states, from the $(70,1^-)$ in the case
of the former and $\frac12^\pm$ in the latter, into the pole diagrams (see
fig.\,\ref{fig:polediag}) used in calculating the radiative decays. On the
other hand, \.Zenczykowski has argued that there are strong indications that
Hara's theorem may indeed be violated. Such a violation would, of course,
imply a failure of one or more of the fundamental input assumptions to the
theorem: gauge-invariance, CP conservation and a local (hadronic) field
theory. The last (in the case of finite-size hadrons) is the weakest of these
assumptions.

The point then is that one cannot infer from this asymmetry alone the
violation (or otherwise) of Hara's theorem. The key to unravelling the
situation can only be found in further experimental data on the other weak
radiative hyperon decays: for example, the experimental asymmetry for the
decay $\XO\to\LO\gamma$ is $0.43\pm0.44$, which, if confirmed as large and
positive, would contradict most of the models that allow Hara's theorem to be
maintained. In any case more data are required to perform serious theoretical
investigations.
\section{NON-LEPTONIC DECAYS}

\subsection{CP Violation}
The subject of CP violation in non-leptonic hyperon decays was addressed in
the talk by Valencia\cite{Valencia:1999x1}. In order to gain access to CP
violation one has to measure asymmetries between hyperon and anti-hyperon
decays. As pointed out by Holstein in his talk, Nature has constructed a
perverse sort of hierarchy, whereby the processes that are easiest to measure
are those least sensitive to CP violation (owing to small prefactors) and
\emph{vice versa}. One of the best candidates in the trade-off between
experimental feasibility and sensitivity is the asymmetry parameter, $\alpha$
(governing the correlation between the parent polarisation and daughter
momentum), in the non-leptonic hyperon decays: \eg, $\LO\to{p}\pi^-$ and
$\XM\to\LO\pi^-$. One thus constructs the following asymmetry:
\begin{eqnarray}
  \mathcal{A}
  &=&
  \frac{\alpha+\bar\alpha}{\alpha-\bar\alpha}
\nonumber \\
  &=&
  - \tan(\delta_P-\delta_S) \, \sin(\phi_P-\phi_S),
\end{eqnarray}
where $\delta_{P,S}$ are the strong ($\Delta{I}=\frac12$) phases and
$\phi_{P,S}$ are the weak (CP-violating) phases. The estimated size of such an
asymmetry, \eg, for the mode $\LO\to{p}\pi^-$ is $\mathcal{O}(10^{-5})$, which
is doable experimentally but tough.

The ingredients that go into the calculation of such an asymmetry are clearly
the two types of phases. The strong phases can be accessed theoretically via
Watson's theorem, which relates $A\to{B}\pi$ to $B\pi$ scattering. This is, of
course, of no practical use for any decay other than $\LO\to{p}\pi^-$. Recent
calculations using chiral perturbation theory suggest that the phases might be
very small for all other modes. For the $S$ and $P$ waves in $\LO$ decay they
are found to be
\begin{eqnarray}
  \delta_S^{\frac12} &\sim&           6^\circ, \\
  \delta_P^{\frac12} &\sim& \llap{$-$}1.1^\circ,
\end{eqnarray}
where the errors are estimated (assumed) to be of the order of $\pm1^\circ$.

The weak phases are calculable via an effective weak-interaction Hamiltonian:
\begin{eqnarray}
\mathcal{H}_\mathrm{eff}^{|\Delta{S}|=1}
  &=&
  \frac{G_\mathrm{F}}{\sqrt2} V^*_{ud}V_{us}
\nonumber \\
  &&
  \mbox{} \times
  \sum_{i=1}^{12} c_i(\mu) \mathcal{O}_i(\mu) + \mathrm{h.c.}
\end{eqnarray}
The short-distance coefficients, $c_i(\mu)$, are well-known while the matrix
elements of the relevant operators are rather model dependent and are
certainly not known with any precision. Using vacuum saturation, one can show
that one of the operators, $\mathcal{O}_6$, dominates; its matrix element is
calculated to be $y_6\simeq-0.08$ and the uncertainty in the calculation
outweighs the uncertainty from neglecting the other operators. At this point
we have
\begin{equation}
  \phi_P-\phi_S \sim -0.4 y_6 A^2\lambda^4\eta,
\end{equation}
where the last three factors are the \CKM matrix parameters of the Wolfenstein
parametrisation and provide $A^2\lambda^4\eta=10^{-3}$; thus, one obtains
$\mathcal{A}(\LO)\simeq-3\cdot10^{-5}$. The errors on such an estimate are
probably best set at around $100\%$.

A particular interest in such numbers is stimulated by the effect of possible
extensions to the Standard Model, the most popular being Supersymmetry. For
example, if the recently confirmed large value for $\epsilon'/\epsilon$ is to
be ascribed to new Supersymmetric couplings, then the same would produced an
enhanced CP-violating asymmetry in non-leptonic decays, with
$\mathcal{A}(\LO)\sim\mathcal{O}(10^{-3})$ being possible.

\subsection{$\Delta{I}=\frac32$ Amplitudes}
A related subject, dealt with by Tandean\cite{Tandean:1999x1}, is the study of
the little-known $\Delta{I}=\frac32$ amplitudes in hyperon non-leptonic
decays. In view of the situation with regard to the $\Delta{I}=\frac12$
amplitudes and the problem of simultaneously fitting the S- and P-wave
contributions, it is instructive to study the $\frac32$ amplitudes. The
analysis presented was based on calculations performed in chiral perturbation
theory\cite{Abdelhady:1999x1}. On the theoretical side the situation is rather
favourable: at leading order, the amplitudes can be described in terms of just
one weak parameter and this can be fixed from the S-wave amplitudes measured
in $\Sigma$ decays. This then allows a full set of predictions for the
P-waves. Unfortunately, as is often the case, the experimental situation is
less favourable and, despite the large corrections found in their one-loop
calculations, the large errors on the measured values does not yet allow a
meaningful comparison of data and theory.
\section{HYPERON POLARISATION}
Another long-standing puzzle in hadronic physics (although a relative
youngster compared to other topics discussed above) is that of the large
transverse hyperon polarisations observed in large-$p_T$ semi-inclusive
hyperon production (for example, see\cite{Bravar:1995fw} for recent data). A
related phenomena is that of the left-right asymmetry in pion production of
transversely polarised targets (see\cite{Adams:1991cs}, for example). The
general phenomenology was presented here by Pondrom\cite{Pondrom:1999x1} and
the more theoretical aspects of the problem were discussed by Soffer in his
talk\cite{Soffer:1999x1}; I would like to enlarge on some of the points made
and touch upon a few others. Let me first stress that, in the absence of
parity violation, the only single-spin asymmetries allowed are those which
correlate the polarisation vector to the normal of the scattering plane, just
as in the examples mentioned above.

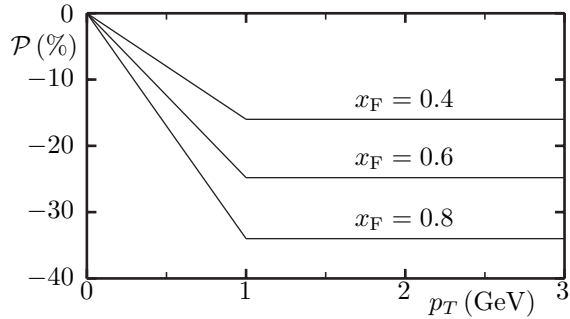
\begin{figure}[htb]
  \centering
\begin{picture}
            (200, 110)(-23,-105)
\LinAxis    (  0,-100)(180,-100)( 3, 2, 3, 0, 1)
\LinAxis    (  0,   0)(180,   0)( 3, 2,-3, 0, 1)
\LinAxis    (  0,-100)(  0,   0)( 4, 1,-3, 0, 1)
\LinAxis    (180,-100)(180,   0)( 4, 1, 3, 0, 1)
\Text       (-16,  -8)[t]{$\mathcal{P}$\,(\%)}
\Text       ( -5,   0)[r]{$  0$}
\Text       ( -5, -25)[r]{$-10$}
\Text       ( -5, -50)[r]{$-20$}
\Text       ( -5, -75)[r]{$-30$}
\Text       ( -5,-100)[r]{$-40$}
\Text       (170,-110)[r]{$p_T$\,(GeV)}
\Text       (  0,-105)[ ]{0}
\Text       ( 60,-105)[ ]{1}
\Text       (120,-105)[ ]{2}
\Text       (180,-105)[ ]{3}
\Line       (  0,   0)( 60, -40)
\Line       (  0,   0)( 60, -62)
\Line       (  0,   0)( 60, -85)
\Line       ( 60, -40)(180, -40)
\Line       ( 60, -62)(180, -62)
\Line       ( 60, -85)(180, -85)
\Text       (120, -33)[c]{$x_\mathrm{F}=0.4$}
\Text       (120, -55)[c]{$x_\mathrm{F}=0.6$}
\Text       (120, -78)[c]{$x_\mathrm{F}=0.8$}
\end{picture}
  \caption{A schematic representation of the semi-inclusive $\LO$ polarisation
    data: the polarisation is given as a function of $p_T$ for various values
    of $x_\mathrm{F}$.}
  \label{fig:lamdapol}
\end{figure}

The archetypal process is $pp\to\LO\strut^\uparrow{X}$, where \emph{neither}
initial-state hadron is polarised while the final-state $\LO$ hyperon is found
to emerge strongly polarised along the normal to the scattering plane. The
principal characteristics of this phenomenon are as follows (see also
Fig.\,\ref{fig:lamdapol}): the polarisation
\begin{enumerate}
\item
is large, reaching values of the order of tens of percent;
\item
grows more-or-less linearly with $x_\mathrm{F}$;
\item
grows more-or-less linearly with $p_T$ up to $p_T\sim1\,$GeV;
\item
remains large and approximately constant for $p_T\gtsim1\,$GeV, up to the
largest measured values of $p_T\sim4-5\,$GeV;
\item
follows the expected SU(6) pattern of signs and relative magnitudes.
\end{enumerate}
To the extent that it has been studied, a similar description also applies to
the pion and other asymmetries where the spin vector belongs to the initial
state.

It is not difficult to see (by expressing the amplitudes in a suitable
helicity basis) that such single-spin asymmetries must be proportional to the
imaginary part of the interference between a spin-flip and a non-flip
amplitude. This poses a two-fold difficulty in gauge theories with light
fermions:
\begin{itemize}
\item
tree-level or Born amplitudes are always real,
\item
spin-flip amplitudes are proportional to a current fermion mass.
\end{itemize}
The first requires loop diagrams, which lead to suppression by a power of
$\alpha_S$ and which typically also lead to colour-factor and kinematical
mismatch. The second, na\"{\i}vely, implies suppression by, at best, a power of the
strange-quark mass divided by $p_T$. Thus, Kane, Pumplin and
Repko\cite{Kane:1978nd} (\emph{prior} to the release of any experimental
results) were led to the conclusion that such large effects would spell doom
for perturbative QCD. As history now tells, the effects were far from zero but
perturbative QCD is still very much alive and kicking!

As usual, there is a ``get-out'' clause: the typical $p_T$ of the data is not
considered large enough yet for perturbative QCD to be reliable. Having said
that, a great deal of progress has been made since the early perturbative
calculations and it is now known that such effects are possible even within a
framework of purely perturbative QCD. Before discussing these developments, I
would like to briefly discuss two of the non-perturbative approaches.

Together with other semi-classical models, Soffer already mentioned the Lund
string-model approach\cite{Andersson:1983ia} and illustrated some of its
shortcomings; I shall add to this by highlighting the inconsistency in the
logic from which it derives such polarisation effects. The initial motivation
is conservation of angular momentum in the string break-up process, producing
the strange anti-strange pair. Orbital angular momentum is generated by a
finite length of string being consumed to produce the energy necessary to
create the pair, which are then necessarily spatially separated. This
separation, combined with a finite $p_T$, leads to non-zero orbital angular
momentum of the pair, which can only be compensated by their spin (\ie, by
aligning or anti-aligning, as necessary). Trivial considerations show that the
predicted sign is correct, assuming the strange quark polarisation is
correlated to that of the final-state hyperon via SU(6) type wave-functions.
However, while the spin of the $s\bar{s}$ pair is limited in magnitude to a
total of one unit, the orbital contribution is essentially unbounded as $p_T$
increases (roughly speaking, $|\vec{L}|\propto{p_T}\sqrt{p_T^2+m^2}$). And
thus the serpent bites its own tail.

Soffer also went into some detail with regard to the use of models based on
Regge theory\cite{Barni:1992qn, Soffer:1992am}. While to a certain degree such
models may provide better insight (they do at least contain explicit reference
to imaginary phases, which the Lund model does not), they cannot expect to
apply to very large $p_T$ configurations. Moreover, all such models are at a
complete loss in trying to explain the large polarisations observed in
anti-hyperon production.

Let me now turn to the developments in perturbative QCD over the past years. A
great deal of new understanding has developed since the days of Kane \emph{et
al.}\ and there are now good reasons for believing that explanations can be
constructed within the framework of perturbative QCD. Nearly fifteen years ago
Efremov and Teryaev pointed out that there exist so-called twist-three
contributions that can come to the rescue\cite{Efremov:1985ip}. First of all,
they note that the mass scale, as required by gauge invariance, is not that of
a current quark but a typical hadronic mass, \ie,
${\sim}\mathcal{O}(1\,\GeV)$.

\begin{figure}[htb]
  \centering
\begin{picture}(150,120)(0,30)
\Line    (  0,141)( 20,141)    \Line    (150,141)(130,141)
\Line    (  0,139)( 20,139)    \Line    (150,139)(130,139)
\Line    ( 11,140)(  8,143)    \Line    (141,140)(138,143)
\Line    ( 11,140)(  8,137)    \Line    (141,140)(138,137)
\Oval    ( 75,140)( 10, 55)(0)
\Line    ( 30,133)( 30,105)    \Line    (120,133)(120,105)
\Line    ( 30,105)(120,105)
\Line    ( 30, 65)(120, 65)
\Line    ( 30, 37)( 30, 65)    \Line    (120, 37)(120, 65)
\Oval    ( 75, 30)( 10, 55)(0)
\Line    (  0, 31)( 20, 31)    \Line    (150, 31)(130, 31)
\Line    (  0, 29)( 20, 29)    \Line    (150, 29)(130, 29)
\Line    ( 11, 30)(  8, 33)    \Line    (141, 30)(138, 33)
\Line    ( 11, 30)(  8, 27)    \Line    (141, 30)(138, 27)
\Gluon   ( 30,105)( 30, 65){-2}{6}
\Gluon   (120,105)(120, 65){ 2}{6}
\Gluon   ( 50,131)( 50,105){ 2}{4}
\DashLine( 75, 10)( 75,160){2}
\Line    ( 37,102)( 43,108)
\Line    ( 43,102)( 37,108)
\end{picture}
  \caption{An example of the twist-three diagrams that may contribute to
    semi-inclusive $\pi$ asymmetry. The dashed line represents the cut through
    the final states, the upper, cut quark line should, in fact, fragments into
    the detected pion. The cross indicates the propagator that reaches the pole
    and that thus provides the imaginary contribution}
  \label{fig:t3diag}
\end{figure}
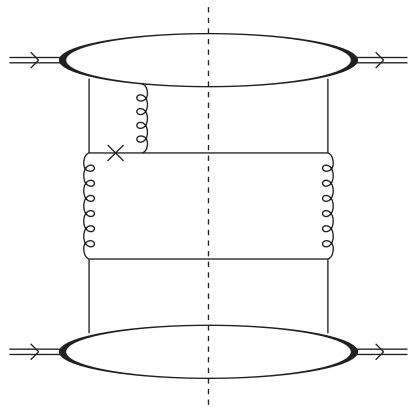

The fact that twist-three contributions are invoked is \emph{not} the
unnecessary complication it might seem. Indeed, it was well-known beforehand
that such would have to be the case owing to the spin-flip requirement:
spin-flip always implies a mass proportionality and therefore higher twist
(note that twist effectively counts the inverse power of $Q^2$, or in this
case $p_T$, that appears in expressions for physical cross-sections). Thus,
the type of diagrams one is led to contemplate are such as that shown in
Fig.\,\ref{fig:t3diag}. The extra gluon leg is attached to the polarised
hadron and is symptomatic of the twist-three nature. The deeper and crucial
observation of Efremov and Teryaev is that when the momentum fraction, $x_g$,
carried by the odd gluon goes to zero, the propagator marked with a cross in
the figure encounters a pole. This is not say that it propagates freely; it is
more a statement of how to perform a contour integral. Indeed, if we adopt the
usual i$\varepsilon$ prescription and split the propagator into its real
(principal value) and imaginary parts:
\begin{equation}
  \frac1{x_gs+\mathrm{i}\varepsilon}
  =
  \mathrm{I\!P} \frac1{x_gs} + \mathrm{i} \pi \delta(x_g),
\end{equation}
where $s$ is the usual (hadronic) Mandelstam variable, then one immediately
sees how an imaginary part arises. Note that, although this may look like a
loop diagram, once again for reasons of gauge invariance, one can show that
the factor of $\alpha_s$ is absorbed into the definition of the hadronic blob
itself and thus it is formally a Born-level contribution. Note also that such
three-legged blobs are exactly what one encounters in the structure function
$g_2$ governing transversely polarised \DIS, presently under study in various
high-energy experiments.

Such diagrams have been exploited by Qiu and Sterman\cite{Qiu:1992wg}, who
have shown that they can produce large asymmetries. Moreover, I have recently
shown\cite{Ratcliffe:1998pq} that in the particular kinematic limit $x_g\to0$
a novel form of factorisation occurs and it is then easy to see why these
diagrams give a contribution of the same order of magnitude as the normal
twist-two Born diagrams.

The advantage of such an approach is that one is clearly free of the usual
model dependence (providing the necessary information on input structure and
fragmentation functions is available). Moreover, through the common
hard-scattering diagrams, such an approach automatically links the many
different possible types of processes in which single-spin asymmetries may be
observed. Note also that it will naturally be applicable to the case of
$\Lambda_c$ polarisation, discussed here by Goldstein\cite{Goldstein:1999x1},
which could then provide a key to the transition from the non-perturbative to
perturbative regimes. On the down side, there are a large number of possible
contributions (\eg, twist-three \emph{fragmentation} functions) and there may
also be a shortage of information, although many phenomenological analyses are
now being performed to identify the origins of the effects and experiments are
continuing to gather information on polarised hadronic structure.
\section{HYPERNUCLEI}
Very little was said at this symposium about hypernuclei. As this is another
long-standing, as-yet unsettled problem and there are experiments planned to
examine it in detail, I decided it would be useful to redress the balance a
little here. The main observation is that a $\LO$ (or even a $\SO$) can move
freely within a nucleus (and indeed nuclear matter in general) without the
usual hinderance of the Pauli exclusion principle, which applies to the
standard nuclear contents: namely, neutrons and protons. Apart from its
slightly larger mass it is very much like (but not \emph{identical} to) a
neutron and therefore is an ideal probe of the nuclear potential and the
forces at work inside a nucleus.

An old question, still debated, is whether or not $\Sigma$-hypernuclei
actually exist. Early reports have never been corroborated although it is hard
to completely rule out the possibility and theoretically there is no solid
argument against their existence.

A more recent and certainly pressing problem regards the observation of an
apparent violation of the age-old $\Delta{I}=\frac12$ rule in the decays of
$\Lambda$-hypernuclei\cite{Bocquet:1986uh}. The situation is, on the face of
it, rather simple. A $\LO$ bound inside a nucleus does not have access to the
standard decay channels $\LO\to{N}\pi$ for reasons of energy. However, it may
decay via an exchange reaction with another nucleon inside the nucleus:
\begin{eqnarray}
  \LO + p & \to & n+p, \\
  \LO + n & \to & n+n.
\end{eqnarray}
From the data on decays of $_\Lambda$H$^4$ $_\Lambda$He$^4$, it appears that
one has the following ratio for the $\Delta{I}=\frac12$ and $\frac32$
amplitudes:
\begin{equation}
  \frac{A(\frac12)}{A(\frac32)} \sim 1.
\end{equation}
It is rather difficult to believe that a nuclear environment with its typical
binding energies of a few MeV could have such a profound effect on hadronic
interactions that have typically much higher energy scales. Indeed, an
interesting approach pioneered by Preparata and
co-workers\cite{Alzetta:1993am}, in which a coherent dynamic pion background
field plays an important r\^{o}le, arrives at results compatible with the
experimental data while avoiding explicit violation of the $\Delta{I}=\frac12$
rule at the hadronic level.

These questions and the nature of the hypernuclei system will come under close
scrutiny in the near future in the FINUDA experiment planned for DA$\Phi$NE
the Frascati $\phi$ factory. This is an $e^+e^-$ facility designed to operate
at a centre-of-mass energy corresponding to the $\phi$ mass (1020\,MeV). The
dominant $K$-$\anti{K}$ decay mode means that this machine will provide a
copious source of $K$ and $\anti{K}$ mesons. In the CHLOE experiment this will
be exploited to study, \eg, CP violation and the $K^0$-$\anti{K}^0$ itself,
and in FINUDA to produce large clean samples of hypernuclei.
\section{CONCLUSIONS}
There is little point in trying to summarise a summary. However, it is worth
making the overall observation that, at least in those areas of hyperon
physics discussed at this symposium, the main stumbling block to progress at
the present is the lack of precision data. Or to put it another way, in many
cases the precision of experiment and theory are roughly equivalent. Indeed,
in some cases the precision of the theory is limited by the lack of precision
input data. This means that it is often impossible to distinguish between the
different models found in the literature and, with no hint as to the way
forward, the theorist is left floundering. During the symposium we have thus
heard various pleas from the theorists for the production of data of better
quality, greater quantity or wider variety, as the case may be, and we have
also heard from the experimentalists that new experiments are planned or in
progress and that new data will be forthcoming.

What is also clear is that many of the problems connected to hyperon physics
are of a very general hadronic nature and thus the answers to the questions
posed here could have far-reaching repercussions. Let us hope that at the next
edition of this symposium some of the models will be swept away, allowing
theorists and experimentalists alike to concentrate their efforts and
resources on the more promising routes.

\end{document}